\documentclass[a4paper,fleqn,usenatbib]{mnras}

\usepackage{newtxtext,newtxmath}
\usepackage{footnote}

\usepackage[T1]{fontenc}
\usepackage{ae,aecompl}

\usepackage{graphicx}	
\usepackage{amsmath}	
\usepackage{amssymb}	
\usepackage{tablefootnote}

\title[]{A (likely) X-ray jet from NGC6217 observed by XMM-Newton}

\author[S. Falocco et al.]{
Serena Falocco,$^{1}$\thanks{E-mail: falocco@kth.se (KTH)}
Josefin Larsson,$^{1}$
Sumana Nandi$^{1}$
\\
$^{1}$KTH, Department of Physics, and the Oskar Klein Centre, AlbaNova, SE 106-91 Stockholm, Sweden\\
}

\date{Accepted 21/08/2017. Received 18/08/2017; in original form 16/06/2017}

\pubyear{2015}

\begin{document}
\label{firstpage}
\pagerange{\pageref{firstpage}--\pageref{lastpage}}
\maketitle

\begin{abstract}
  NGC6217 is a nearby spiral galaxy with a starburst region near its
  center. Evidence for a low luminosity Active Galactic Nucleus (AGN) in its core
  has also been found in optical spectra. Intriguingly, X-ray observations by ROSAT revealed three knots aligned with the galaxy
  center, resembling a jet
  structure.   This paper presents a study of XMM-Newton observations made to assess the hypothesis
  of a jet emitted from the center of NGC6217.  The XMM data confirm the knots found with
  ROSAT and our spectral analysis shows that they have similar spectral properties with a hard photon index $\Gamma\sim1.7$. The
  core of NGC6217 is well fitted by a model with an AGN and a
  starburst component, where the AGN contributes at most 46\% of the total flux.
  The candidate jet has an apparent length $\sim15$ kpc and a luminosity of $\sim5\times10^{38}$~erg $\rm{s}^{-1}$.
It stands out by being hosted by a spiral galaxy, since jets are more widely associated with
  ellipticals.
   To explain the jet launching mechanism we consider the
  hypothesis of an advection dominated accretion flow with a low
  accretion rate. The candidate jet emitted from NGC6217 is intriguing
  since it represents a challenge to the current knowledge of the connection between AGN, jets and host galaxies.
\end{abstract}




\section{Introduction}
In the last two decades the census of X-ray jets detected in Active Galactic Nuclei (AGN) has improved significantly \citep{harris2006,massaro2011,worrall2009,birkinshaw2002,kraft2005,worrall2012,hardcastle2007,hardcastle2016}. Despite the huge progress in understanding these objects, unresolved issues still remain; for instance, it is not clear yet why only a small percentage of AGN display jets or how they form.
Jets can be extended up to $\sim$ 100 kpc, thus a fundamental question involves the collimation mechanism over such extended spatial scales,  bigger than the typical host galaxy scales \citep{bagchi2007,hocuk2010}.
Jets and outflows injecting material far away from the nucleus may play a crucial role in AGN feedback, the self-regulating mechanism preventing the further growth of Super Massive Black Holes (SMBH).
Investigating the connection between jets, AGN activity and host galaxies is the key to achieving a comprehensive picture of these sources.

NGC6217 is a nearby (z=0.0045) barred spiral galaxy with an inclination of $37.7^{\circ}$ \citep{cabrera2004}. 
Interestingly, NGC6217 displays signatures of a large-scale X-ray jet, which is uncommon for spiral galaxies \citep{mao2015} although there are now some confirmed examples (e.g. \citealt{ledlow2001,croston2008,mingo2011,irwin2017}).
The putative jet, discovered with ROSAT, has a discrete structure formed by three knots directed South-West (SW) from the core \citep{pietsch2001}. The apparent jet length is 15 kpc and the total jet luminosity was reported to be $1.7\times 10^{39}$~erg $\rm{s}^{-1}$,  while the luminosity of the central core was $10^{40}$~erg $\rm{s}^{-1}$.
For NGC6217, the jet proposed by \cite{pietsch2001} would require an AGN nucleus since starbursts are not likely to form such collimated structures. The AGN may be accreting through an Advection Dominated Accretion Flow (ADAF), as is found for a number of radio galaxies with long relativistic jets  (e.g. \citealt{hardcastle2006,hardcastle2007a,hardcastle2009}).  The identification of the nature of the core of NGC6217 is of primary importance to provide support to or reject the jet hypothesis.


NGC6217 has been classified as a Low Ionisation Nuclear Emission Line Region (LINER) on the basis of its optical spectrum \citep{nicholson1997}.
The energy source that excites the line emitting gas in LINERs is still debated since some sources display features of faint AGN, while others are instead pure starbursts. 
The X-ray variability of NGC6217 has been investigated as a proxy for AGN activity by \cite{nicholson1997}, using ROSAT data. No variability was detected, so an AGN core is unlikely to dominate the observed X-ray emission. \cite{cabrera2004} classified the galaxy as having a Seyfert 2 nucleus. There is also evidence of a starburst region located $10''$ from the nucleus along the bar to the south east \citep{artamonov1999}, as well as 
irregular dust lanes and patches passing through or close to the centre \citep{malkan1998,rutkowski2013}.

In this paper we present an analysis of XMM-Newton images and spectra of NGC6217 and its putative jet.  Section 2 describes the observations and data reduction, Section 3 reports the analysis and results, Section 4 discusses the main findings on the basis of a comparison with previous studies, and Section 5 presents the conclusions and future perspectives. 
The
spectra have been analysed with {\sc Xspec} version 12.9. The errors
correspond to a 90\% confidence level for one interesting parameter ($\Delta\chi^2$ = 2.71) unless otherwise stated. The cosmological
parameters are the default ones in {\sc Xspec}: $H_0=70\  \rm{km/s/Mpc}$, $\Omega_\Lambda=0.73$, $\Omega_{\rm{tot}}=1$.

\section{Observations and data reduction}

The main focus of our analysis are the X-ray observations of NGC6217 by XMM-Newton, described in \ref{xmmobs} below.  For comparison, we also consider archival optical (Hubble Space Telescope, HST) and radio (Very Large Array, VLA) observations. These  are described in sections \ref{hstobs} and \ref{radioobs}, respectively.

\subsection{XMM-Newton observations}
\label{xmmobs}
Four observations from XMM-Newton are available for NGC6217,  they are listed in Table \ref{t:obs}. All the observations were performed in Prime Full Window mode. All the MOS observations have the medium filter, while the pn observations have the thin filter. The background was stable and did not display flares in any of the observations. In the first observation only MOS2 data were used since the core of NGC6217 fell along a CCD gap in MOS1 and the pn was not active.  The data were reduced using {\sc SAS} (Science Analysis System\footnote{'Users Guide to the XMM-Newton Science Analysis System', Issue 13.0, 2017 (ESA: XMM-Newton SOC)}) version 15.0.0 of February 2016 and the CCF files are XMM-CCF-REL-334 (from 12 May 2016). Images and spectra of the nucleus and the knots were extracted as described below.
 Optical Monitor (OM) observations were carried out in the imaging mode during all four observations. The UV filters $uvw1,uvm2$  and $uvw2$ were used in all observations, while the two observation in 2007 also included the $u$ filter besides the UV filters.  We reduced these data using the {\sc SAS} task {\sc omichain}.

\begin{table}
  \begin{tabular}{c c c c}
    OBS ID & date & exposure (s) & instrument \\
     (1) & (2) & (3)   & (4)  \\
    0061940301  & 2001-09-20 & 6135  & MOS2 \tablefootnote{The MOS1 observation has not been used for the core because it is along a CCD gap; the pn is not active}  \\
 0061940901  &   2002-04-11 & 9713 & MOS1/MOS2/PN \\
 0400920101   &  2007-02-15  & 40374 & MOS1/MOS2/PN\\
 0400920201  & 2007-02-17 & 38869  & MOS1/MOS2/PN\\
  \end{tabular}
  \caption{XMM-Newton observations of NGC6217.  (1): obs-id as defined in the XMM catalogue; (2): observation date; (3): exposure time; (4): instruments used for the spectral analysis}
   \label{t:obs}
\end{table}

\begin{table}
  \begin{tabular}{c c c c c}
    ra & dec &  Name &  HR \tablefootnote{Hardness Ratio defined as (H-S)/(H+S) where H are the counts in 2-10 keV and S in 0.3-2 keV (observed frame). } \\
    (1) & (2) & (3)   & (4)  \\
    16:32:38.79   & +78:11:51.1    &  NGC6217  &   $-0.93\pm0.12$ \\
      16:32:25.44  &   +78:11:18.2     &    knot1 &  $-0.42\pm0.14$ \\
  16:32:13.50  &  +78:10:29.7    &  knot2   &  $-0.56\pm0.15$  \\
 16:31:59.88   &  +78:09:58.2    & knot3 &   $-0.46\pm0.14$ \\
  \end{tabular}
  \caption{(1) and (2): source coordinates, (3): identification, (4): hardness ratios from XMM-Newton data} 
  \label{t:sources}
\end{table}

\subsubsection{Imaging}
The XMM observations of each epoch and each camera were aligned and merged in order to improve the signal-to-noise. 
The alignment code makes use of four point sources in the field, located away from chip gaps and borders. The choice of the specific sources does not significantly affect the final merged image. The aligned images from MOS1, MOS2 and pn were combined through the straight sum in each pixel. We analyse the images in the soft band, between 0.3 and 2 keV observed frame, due to the low number of counts at higher energies.

    \begin{figure}
   \includegraphics[width=\linewidth,height=7cm]{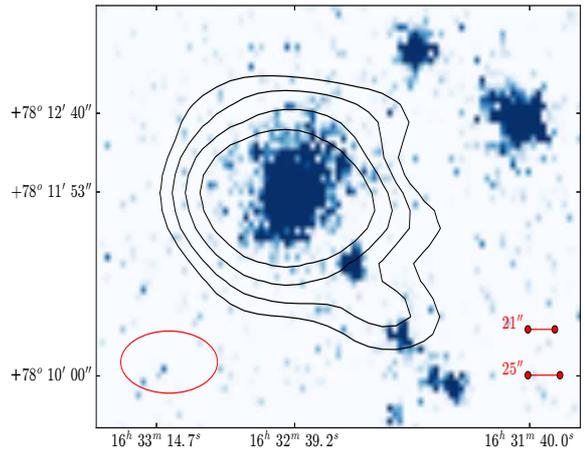}
 \caption{ Stacked XMM-Newton image of NGC6217 from the multiple exposures of MOS1, MOS2 and pn in the $0.3-2$~keV  band. The overplotted contours are from the VLA archival 1.4 GHz image of 1984. The radio contours are 0.001, 0.002, 0.004 and 0.008 $\rm{mJy~beam^{-1}}$.  FOV: 354 '' x 312 ''. Bottom left: beam of the radio image. Bottom right: radii of the extraction regions for the knots (top) and the source (bottom).}
  \label{fig:xmm}
  \end{figure}

    \begin{figure}
   \includegraphics[width=7cm,height=7cm]{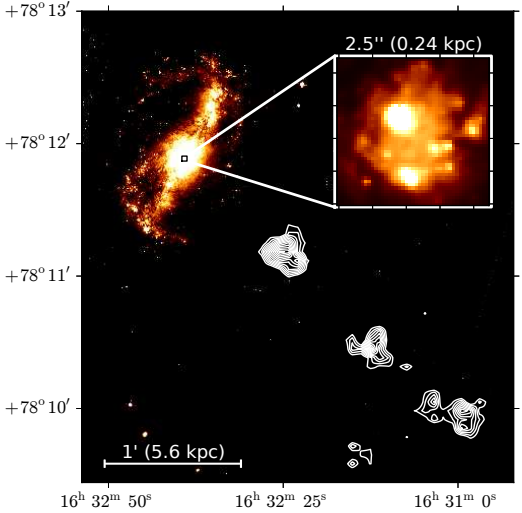}
   \caption{HST ACS/F625W image (colour) of NGC6217 together with a contour map of the X-ray sources to the SW (black and white, from the stacked 0.3-2 keV XMM image, Fig. \ref{fig:xmm}). FOV: 190 '' x 213 ''. The inset shows a zoom of the central core of NGC6217. 
   }
  \label{fig:hst_xmm}
  \end{figure}

    \begin{figure}
   \includegraphics[width=\linewidth]{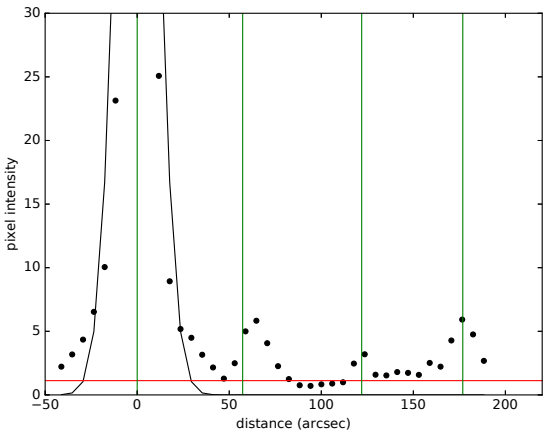}
   \caption{Radial profiles in the direction of the knots, from the XMM image in Fig. \ref{fig:xmm} (average intensity over 3x3 pixel boxes). The red line marks the background level. The black line is a Gaussian fit to a point source in the field. The peak of the central source is 75 counts/pixels.   Vertical lines mark the ROSAT positions of the knots.}
   \label{fig:xmm_profiles}
\end{figure}

 \subsubsection{Spectra}
\label{xmmspecred}
The spectra of the nucleus and the knots were extracted in circular regions centered at the positions of the sources, with radii of 25 arcsec for the nucleus and 21 arcsec for the knots. The background spectra were extracted from source-free areas falling in the same chips as the source.
For each observation, the spectra from MOS1 and MOS2 were merged by computing the straight sum of the counts and the exposures. The backscale of the total spectrum was computed as the average of the backscale values of the individual spectra, weighted by the contribution of each exposure. The response matrix and ancillary file of the combined spectrum were constructed in the same way, using the weighted average  of the individual ones (computed using the ftools task addrmf). 

The same procedure was applied to also merge observations from nearby epochs, which have the same filter and similar off-axis angles. 
This procedure resulted in four spectra; one pn spectrum from 2002, one pn merged from the two observations in 2007, one MOS merged from 2001 and 2002, and one MOS merged from the observations in 2007.  We applied the same merging procedure for the spectra of each of the three knots.

\subsection{HST observations}
\label{hstobs}

NGC6217 has been observed by HST multiple times with different cameras and filters. We selected a recent, high-quality observation in order to compare the optical properties of the host galaxy with the XMM-Newton image. Specifically, we used the ACS/F625W observation from 2009-06-13. The image was created using {\sc DrizzlePac}\footnote{http://drizzlepac.stsci.edu}, following standard procedures.

\subsection{Radio observations}
\label{radioobs}

There are a number of archival VLA images available for NGC6217: 1.49~GHz (L-band) from 1982 and 1984; 4.89~GHz (C-band) from 1985 and 1988; 8.49~GHz (X-band) from 2001.  We used the FITS files generated by the VLA pipeline in AIPS and available from the VLA archive pilot page.

\begin{table*}
  \caption[]{Simultaneous fits of MOS and PN spectra. ME=wa*zwa*mek,  MEME= wa*(zwa*mek+zwa*mek);
    PL=wa*zwa*pow;
    MEPL: wa*(zwa*pow+zwa*mekal);
    (1): model used to fit the spectrum in the 0.5-2.5  keV rest frame band; (2): $\rm{N_H}$ of the cold mekal component (or of  the powerlaw in PL and MEPL models);  (3): $\rm{N_H}$ of the hot mekal component (or of the only mekal in MEPL model); (4): total observed flux in the full band (not corrected for absorption) in the 0.5 - 2.5 keV rest-frame band; (5): temperature of the 'cold' thermal emission; 
    (6): Temperature of the 'hot' thermal emission; (7):  chi2/dof; (8): observed flux of the 'cold' component in MEME model (or flux of the only thermal component in MEPL model) in the 0.5 - 2.5 keV rest-frame band (not corrected for the absorption); (9): observed flux of the 'hot' thermal component (or flux of the powerlaw in MEPL model) in the 0.5 - 2.5 keV rest-frame band (not corrected for the absorption); (10): knot identification. 'knots' stands for a simultaneous fit of the three knots. (11) The model applied to the knot spectra.
    }

\begin{tabular}{c c c c c c c c c c}       

  model  & $\rm{N_{H,c}}$ & $\rm{N_{H,h}}$  & F  & $\rm{T_c}$ & $\rm{T_h}$  & $\chi^2$/dof   & $\rm{F_c}$ &  $\rm{F_h}$ \\
       (1) & (2) & (3) & (4)  & (5) & (6)  & (7) &  (8) &(9)  \\
    & ($10^{22}\rm{cm^{-2}}$) & ($10^{22}\rm{cm^{-2}}$)  & ($10^{-13}\rm{erg}$ $\rm{cm^{-2} s^{-1}}$)  & (keV) & (keV)  &    & ($10^{-14}\rm{erg}$ $\rm{cm^{-2} s^{-1}}$) &  ($10^{-14}\rm{erg}$ $\rm{cm^{-2} s^{-1}}$) \\\\

  ME    &  $0.25\pm0.05$  & -    &  -  & -  & $0.597\pm0.019$  &    431.00/180    & -   & - \\  \\
  MEME    & $<0.074$  &   $0.61\pm0.07$  & $1.61\pm0.37$    &  $0.30_{-0.05}^{+0.02}$ & $0.67_{-0.04}^{+0.07}$  &     187.56/177    &  $6.67\pm1.67$   &  $9.37\pm1.78$ \\  \\


  \hline
   model  &$\rm{N_{H,pow}}$ & $\rm{N_{H,th}}$  & F  & $\Gamma$ & T  & $\chi^2/dof$   & $\rm{F_{pow}}$  & $\rm{F_{th}}$ \\
       & ($10^{22}\rm{cm^{-2}}$) & ($10^{22}\rm{cm^{-2}}$)  & ($10^{-13}\rm{erg}$ $\rm{cm^{-2} s^{-1}}$)  &  & (keV)  &    & ($10^{-14}\rm{erg}$ $\rm{cm^{-2} s^{-1}}$) &  ($10^{-14}\rm{erg}$ $\rm{cm^{-2} s^{-1}}$) \\
   \\
   PL    &  $0.62\pm0.08$  & -  &   $1.51\pm0.32$
  & $6.6\pm0.5$      & -  &     348.35/183  & -  & -    \\  \\
   MEPL  &   $ <0.179$   &  $ 0.15^{+0.11}_{-0.14}$    &  $1.61\pm0.56$  &   $2.95^{+0.81}_{-0.34}$      & $0.62\pm0.03$  &     186.31/177   &   $ 7.42\pm2.00$ &    $8.64\pm3.89 $\\  \\


       \hline \hline
       source  &  model  &  & F  & $\Gamma$  & & c-stat/dof   &  \\
       (10)  &  (11)  &  &  &  &   & &    \\  \\
       &    &  & ($\rm 10^{-15}\rm{erg}$ $\rm{cm^{-2} s^{-1}}$) &  &   & &    \\  \\

       knot1       & PL  &        &   $12\pm7$  &  $1.74\pm0.47$  &-  &     155.64/192     &     &   \\  \\



knot2       & PL  &          &   $9\pm3$    &  $1.71\pm0.53$    &-  &        128.23/165 &     &  \\  \\



knot3       & PL &        &  $ 12\pm4$   &   $   1.70\pm0.46 $  &- &       161.5/210  &    &   \\  \\



knots       &PL  &        & $ 11\pm7$   &  $   1.72\pm0.28 $  &- &      445.4/569  &    &   \\  \\

knots      & ME  &            & $11\pm7 $  &   & $5^{+5}_{-2}$  &   446.29/569   &     &   \\  \\

\end{tabular}
\label{t:fits}
\end{table*}
  
\section{Analysis and results}
We used two independent approaches to analyse the X-ray data available for NGC6217: imaging and spectroscopy. The sections below describe the results obtained with the two approaches.
\subsection{Images}

The soft X-ray image resulting from the alignment and merging is shown in figure \ref{fig:xmm}, where the radio contours (VLA archival 1.4 GHz image of 1984) are over-plotted. The radio data will be discussed at the end of this section. 

The X-ray image shows a structure formed by three knots directed South West (SW) from the nucleus of NGC6217.
It is directed in an apparent perpendicular direction to the disk of the galaxy, as shown in Fig. \ref{fig:hst_xmm}, where the optical image from HST has been over-plotted with the soft X-ray contours. The HST image shows a complex structure in the galaxy centre, formed by two main nuclei, 1.3 '' apart, surrounded by fainter knots and diffuse emission (see Fig. \ref{fig:hst_xmm}). This structure cannot be resolved with XMM-Newton, which has a half-energy radius significantly larger than the size of the structure (15'').

The putative jet is extended over 160'' (15~kpc assuming redshift z=0.0045) from the XMM-Newton images, consistent with the ROSAT results. The two sources seen in the North West direction from NGC6217 have been detected by ROSAT and they are source 68 and 74 in the  catalogue presented in \citep{pietsch2001}. Both sources are AGN, with a redshift of 0.38 for source 68, but no redshift reported for source 74.  
We have computed the hardness ratios (HR) of the nucleus of NGC6217 and the knots using the counts in the observed frame bands 0.3-2 keV and 2-10 keV to obtain information about the broad band X-ray colours. As reported in Table \ref{t:sources}, we found that the core and the knots are mainly soft-X-ray emitters and that the three knots have a higher HR than the core. This is also confirmed by the spectral analysis detailed below. 
The X-ray colours and the more accurate estimates of the photon index of the knots are consistent within the uncertainties with those of other jets in the literature (e.g. \citealt{kraft2005,hardcastle2009,worrall2016}).  

Fig. \ref{fig:xmm_profiles} shows the radial profiles
from the soft image as the pixel intensity in the direction of the knots, averaged in boxes of $3\times3$ pixels. The red line is the background
level computed as the average pixel intensity plus one sigma in a stripe parallel to the core-knots
direction (124.8 arcseconds below it) and free from bright sources. The specific choice of the stripe to compute the background level does not change the results significantly. The first and third knot are well above the background level, while
the second knot is weaker. This is consistent with the results obtained with ROSAT, where the second knot was the faintest one \citep{pietsch2001}. The black, solid line is a Gaussian fit to a point-source profile in the field (with $\sigma\sim$10'').
The positions of the knots in the XMM images are consistent with the ones found in the ROSAT data, represented in green in Fig. \ref{fig:xmm_profiles}, within the positional errors introduced by the instrumentation. 

The excess between the observed profile of NGC6217 and the Gaussian provides us with a hint
for a marginal extension, but the radial profile in a different direction (perpendicular to the knots) does not support this evidence, showing instead a profile consistent with a point-like source. The knots do not show an extended profile.

We also compared the X-ray and optical images with archival radio data obtained by VLA  at 1.49 GHz, 4.89~GHz and 8.49~GHz with different array configurations. The radio images have been shown in \cite{hummel1984}, \cite{condon1987}, \cite{pietsch2001}, \cite{vila1990}  and discussed  in \cite{saikia1994,saikia1995}. The low-resolution images at 1.4 GHz show a single component with a weak extended emission along the south west direction  
(see contours overlaid on the XMM-Newton image in Fig. \ref{fig:xmm}). No correspondence to the X-ray knots is seen, but we note that the images have poor sensitivity and that these data are not adequate to resolve such compact scales.  A more sensitive high resolution observation is required to confirm the presence of jet. 
The other available radio images only probe the central region of the galaxy. In particular, the intermediate resolution maps ($\sim$2 arcsec) at 4.8 GHz B-array and 8 GHz C-array show an extended structure in the North West (NW) direction and a bright emission peak. The high resolution ($\sim$0.64 arcsec) 4.8 GHz A-array image of the nuclear region reveals a complex extended structure in the NW direction and a bright bending structure towards the south (peak flux 1.08 mjy/beam at RA = 16:32:39.54811, Dec = 78:11:53.3246). 
This extra-nuclear region of complex condensations may be driven by a starburst \citep{hummel1984}. More sensitive VLA A-array observations are required to constrain the spectral index distribution for this region. The estimated integrated spectral index using all the available VLA maps is $\sim$ 1. This spectral index indicates that the source is dominated by non-thermal emission. The limited absolute astrometry of HST prevents us from drawing any conclusion about correlations between the optical and radio images in the nuclear region.

\subsection{Spectra}
\label{spectra}
The spectral analysis of the core was carried out with the purpose of finding its best-fit model, discriminating between different physical pictures. Because of limited statistics, the spectra of the knots have instead been analysed   with the main purpose to determine the spectral slope and to compare the spectra of the individual knots. 
The spectra of the core were grouped in order to have 15 counts per bin and then fitted using $\chi^2$ statistics. Instead, the spectra of the knots, since they have less counts, have been binned with 1 count in each bin and fitted using Cash statistics.
\subsubsection{Spectra of the core}

The core spectra from the four epochs were first fitted individually to check for flux and spectral variability. The first observation in 2001 had too low count rate to yield any strong constraints. For the other three epochs, we found that the flux is constant within the uncertainties:  $(1.7\pm1.0)\times10^{13}~\rm erg$ $\rm{cm^{-2}\ s^{-1}}$ in 2002, $(1.5\pm0.4)\times10^{13}~\rm erg$ $\rm{cm^{-2}\ s^{-1}}$ for the first observation of 2007, and $(1.5\pm0.4)\times10^{13}~\rm erg$ $\rm{cm^{-2}\ s^{-1}}$ for the 2nd observation in 2007. The spectra did not display any significant spectral variability.
 We therefore merged the spectra from nearby epochs (see section \ref{xmmspecred}) and simultaneously fitted the merged spectra to increase the count statistics. The  fits were performed with a free constant of normalisation between epochs and detectors, but all other parameters tied between the spectra. The spectra were fitted in the rest-frame band between 0.5~keV and 2.5~keV. We did not extend our study above 2.5~keV because of the low number of counts at these energies. The channels below 0.5 keV were excluded due to the high instrumental background in XMM-Newton.

 The amount of Galactic absorption in the direction of NGC6217 is 3.91 $\times10^{20}\rm cm^{-2}$. This value has been fixed during the spectral fitting.
For the core we consider two scenarios: (1) the core is a pure starburst; (2) the core hosts a starburst region and an AGN nucleus. Within scenario (1), we use the following models:
\begin{itemize}
\item a thermal component with Bremmstrahlug emission and collisional lines (\emph{mekal} in {\sc xspec}), used to represent the emission from the starburst region. The abundances and densities are fixed to the default values in {\sc Xspec}. The only free parameters in the \emph{mekal} model are thus the temperature of the plasma and the normalisation. The model is absorbed with photoelectric absorption in the source itself in addition to the Galactic absorption. The intrinsic absorption is free to vary. This model is called ME in Table \ref{t:fits}.
\item two thermal emission components absorbed by different column densities (both free to vary during the fitting). As above, only the  temperatures and the normalisations are free to vary in the \emph{mekal} models. This model takes into account the starburst nucleus and any thermal emission from the spiral arms. The model is called MEME in Table \ref{t:fits}.
\end{itemize}
As far as concerning the hypothesis of the existence of an AGN nucleus, we use the following models:
\begin{itemize}
  \item  an  absorbed powerlaw  to represent the AGN continuum (free intrinsic column density, spectral slope and normalisation). We use a simple powerlaw to represent the AGN because there is no evidence for reflection features such as iron lines or Compton reflection. This model is called PL in Table \ref{t:fits}
  \item  a powerlaw and a thermal emission component (\emph{mekal}), absorbed by different column densities. The free parameters are the same as described above. This model is called MEPL in Table \ref{t:fits}.
\end{itemize}

It would be natural to also consider a combination of these two scenarios, i.e. a model with two thermal components as well as an AGN contribution. However, we do not have enough statistics to constrain the parameters of such a model. 

All the fit results are presented in Table \ref{t:fits}.
From the $\chi^2$ in scenario (1), it is clear that a single \emph{mekal} is not enough to fit the data and that a second thermal component is required. Similarly, in scenario (2), a simple powerlaw fails to fit the data, with a bad $\chi^2$ and a non-physical value of the photon index. The spectra are better fitted with a second component of thermal emission. With the addition of the second thermal component in scenario (1), the data are well fitted with two thermal plasmas, one at a temperature of $\sim$0.3 keV and one at $\sim$0.7 keV. The colder thermal component can be explained as emission from the spiral arms, while the hotter component is most likely from a nuclear starburst (as in M83 \citep{soria2002} and NGC253 \citep{pietsch2001b}). The fit with model (2) finds a
nuclear starburst (T$\sim$0.6 keV), and a steep power law   ($\Gamma\sim$ 3 $\pm$ 0.5). This photon index is consistent with the X-ray spectra of radio-loud sources where the jet contributes to the soft X-ray spectra \citep{hardcastle2007,hardcastle2009,mingo2014}.  

\begin{figure*}
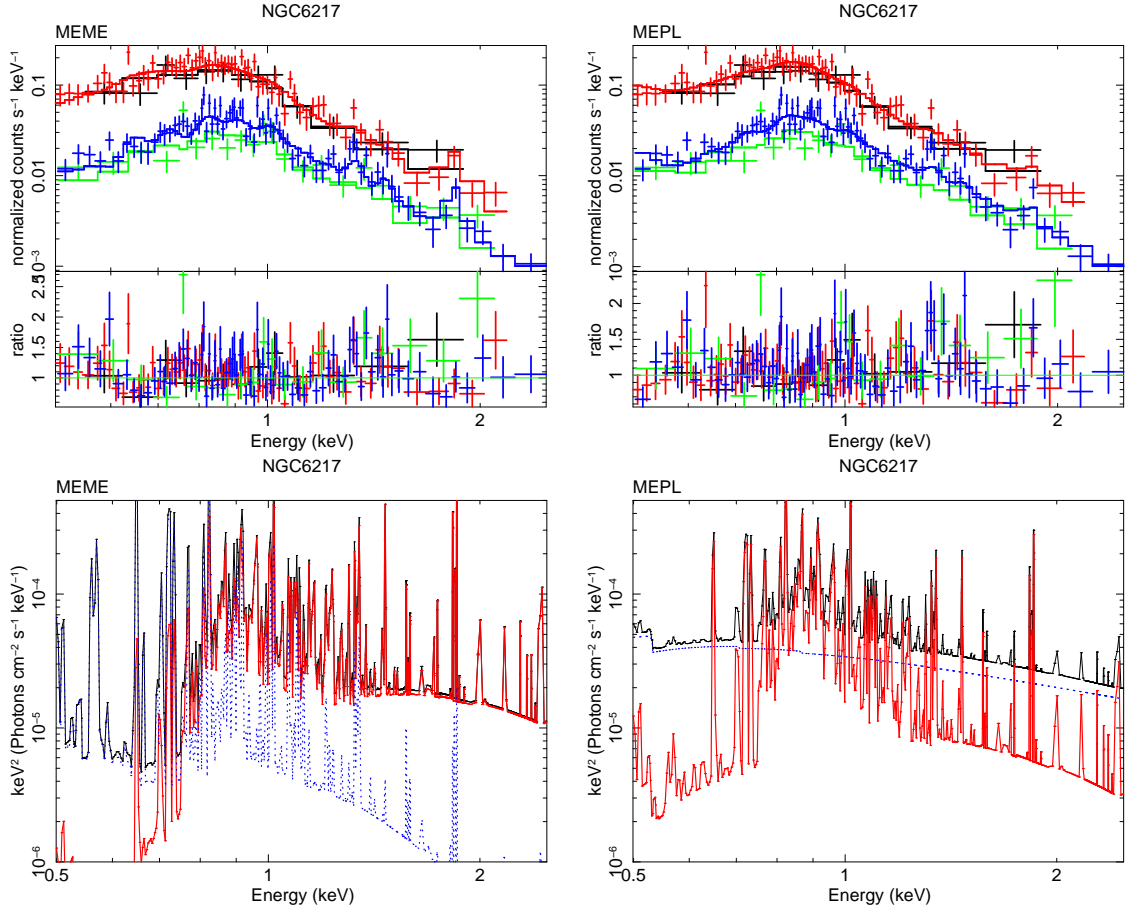

   \centering
    \includegraphics[width=6cm,angle=-90]{figures/const_wa_zwa_mek_zwa_mek_pn_mos_source.ps}
    \includegraphics[width=6cm,angle=-90]{figures/const_wa_zwa_pow_zwa_mekal_pn_mos_source.ps}\\
    \includegraphics[width=6cm,angle=-90]{figures/const_wa_zwa_mek_zwa_mek_pn_mos_source_one_spectrum_eemodel.ps}
      \includegraphics[width=6cm,angle=-90]{figures/const_wa_zwa_pow_zwa_mekal_pn_mos_source_one_spectrum_eemodel.ps}
\caption{Spectra of the core of NGC6217 fitted with different models. { \emph{Top-left panel}: Fit with MEME (see Table \ref{t:fits}) and ratio between the data and model. }\emph{Top-right panel:} Fit with MEPL (see Table \ref{t:fits}) and ratio between the data and model.
Black:  pn observation of 2002; Red: pn observations of 2007 (merged); Green: MOS spectra of 2001 and 2002 (merged); Blue: MOS spectra of 2007 (merged).
 \emph{Bottom left panel}: Components of the MEME model. Red: starburst component; Blue: spiral arms contribution.
\emph{Bottom right panel}: Components of the MEPL model. Red: starburst component; Blue: powerlaw (AGN) component. The black line in the bottom plots shows the resulting model.
  }
          \label{fits}
    \end{figure*}

Fig. \ref{fits} shows
the spectra fitted with the models MEME and MEPL in the scenarios (1) and (2), respectively, together with the corresponding ratios between the data and the model.
The different normalisation of the MOS and pn spectra are due to their different instrumental responses, while the better spectral quality in the spectra of 2007 compared to 2001 and 2002 are due to longer exposures (see Table \ref{t:obs}).
It is apparent from the figures that both models fit the observed spectra equally well. The two models are displayed in the lower panels of Fig. \ref{fits}.
It is clear that the main difference between the models is at energies below 0.7 keV, where there is a contribution from the arms or the AGN continuum, depending on the model. In the intermediate energy range ($0.7$ keV$<E<1.5$ keV), both the additive models that compose MEME and MEPL contribute about equally to the observed spectra. In the high energy range ($E>1.5$ keV), the spectrum has a major contribution from the powerlaw emission (AGN), with some lines from the starburst, when fitted with the MEPL model. Alternatively, it is also well represented by the high temperature \emph{mekal} (the starburst) if the MEME model is used, where the continuum is due to Bremmstrahlug. 

Our fits show that, while a hot starburst component is required in all fits within both the pure starburst and the starburst-AGN scenario, the current data do not allow us to pinpoint the AGN continuum. The total X-ray flux is 1.5$\sim10^{-13} \rm{erg}$ $\rm{cm^{-2} s^{-1}}$, out of which the powerlaw component contributes   46\%  in the MEPL model. This serves as an upper limit on the possible AGN contribution. The constrains on the intrinsic column densities obtained from the fits show that the nucleus of NGC6217 is not strongly absorbed.

\subsubsection{Spectra of the knots}

We fitted the spectra of the knots with a powerlaw model (PL in Table \ref{t:fits}). The knots have spectral properties consistent between them, within $30 \%$ uncertainties. Interestingly, they are also harder than the main source ($\Gamma\sim$ 1.7), confirming the results based on the HR study.  
We also fitted the three knots simultaneously since they have compatible spectra, and obtained better constraints on their spectral slope, resulting in  $\Gamma$=1.72$\pm$0.28. To investigate a possible thermal origin of the spectra we also fitted a {\it mekal} model (ME in Table \ref{t:fits}), but the fits to the individual knots leave the temperatures unconstrained. The joint fit of the three knots results in a high temperature between 2 keV and 10 keV at 90 \% confidence level. 

\section{Discussion}
Imaging and X-ray spectral analysis of the XMM observations of NGC6217 have been performed to investigate the nature of the core and the knots aligned with it. The following section discusses the main findings, comparing them with examples of extra-galactic jets in the literature and with previous studies of NGC6217.
\subsection{Jet of NGC6217}

The XMM images display a jet structure formed by three knots aligned with the core of NGC6217. For the interpretation of the three knots, scenarios alternative to the jet are worth discussing. These include the hypothesis that the three knots
are unrelated background sources or that they instead compose a starburst-driven super-wind.

\subsubsection{Background sources scenario}
The
 scenario where the knots are due to unrelated background sources is unlikely if the spectra are consistent with each other, because it would imply that three very similar sources are aligned with the nucleus.
The XMM data studied in this work reveal similarities between the knots: they have similar spectra and X-ray colours (they are soft). Observations at other wavelengths, for example in the optical and radio bands, would place stronger constraints on the nature of the knots.  The HST data show no clear optical counterparts of the knots, as reported for other well known X-ray jets in the literature (e.g. Cen A, \citealt{dejong2008} as well as those in \citealt{sambruna2004}). The knots are not detected in archival radio images either, although we note that these are of limited quality.

To place stronger constraints on the possibility that the knots are unrelated background sources (most likely AGN), we searched for additional optical information and also compared the results with the two nearby background AGN reported in \cite{pietsch2001} (see section 3.1). We used the information extracted from the simultaneous OM data and searched for associations within a radius of 2 arcsec around the sources of interest. Associations were detected only in the OM $u$ filter, so all the magnitudes reported refer to this band. For the background AGN 74, \cite{pietsch2001} reported an optical magnitude of 18.4 and an  X-ray to optical flux ratio of -0.6. There is an association with an optical source in the two OM observations from 2007 with magnitudes 19.4$\pm0.2$ and 20.05$\pm0.10$, respectively. The background AGN 68 has an optical magnitude 19.7 and an X-ray to optical flux ratio of -0.3 from \cite{pietsch2001}. It has counterparts in the OM data of 2002, and in both 2007 observations, with magnitudes 17.8$\pm0.2$, 17.70$\pm0.09$, 18.19$\pm0.04$, respectively. 

To search for optical association with the three knots, we considered both the OM data and the Sloan Digital Sky Survey (SDSS). The results from this can be summarised as follows:

\begin{itemize}

\item Knot 1. There are no associations with OM or SDSS sources. Assuming that the source is an AGN, its optical and X-ray fluxes would be inside a region mostly populated by AGN defined as: $ -1 < \log{\frac{F_x}{F_o}}<1  $ \citep{mainieri2002}. Extrapolating the optical flux from the X-ray flux reported in Table 3, the optical magnitude would be between 20.8 and 25.8. The source may therefore be detectable in the OM, given that the faintest sources detected in these observations have magnitude 21.6. Moreover, the optical flux estimates are well above the sensitivity limits of the SDSS. An optical counterpart would therefore have been detected if the source was an AGN. 

\item Knot 2. There is one association at 1.6 arcsec from knot 2 in the last 2007 OM observation. Its $u$ magnitude is 20.3$\pm0.1$. There is also a source at 0.9 arcsec from knot 2 in SDSS with r magnitude 20.3. The SDSS source is consistent with the OM source, despite their different distances from the X-ray knot, considering the positional uncertainties of the XMM data. The SDSS source has been classified as a stellar object and its colours are: g-r=0.11, r-i=-0.008,  i-z=0.18. A very faint point source consistent with this position is also is seen in the HST image. Given its colours and its point-like profile, this source can be interpreted as an unrelated star.

\item Knot 3. There are no associations in any of the OM observations. As for knot 1, assuming that the source was an AGN its optical magnitude would be  between 20.8 and 25.8, and it would be possible to detect it in the OM data. The search for associations in SDSS revealed an extended source 1.2 arcsec away from the knot. Its magnitude in the r band is 21.5 and its optical colours are: g-r=0.063, r-i=0.4, i-z=-0.019. Inspecting the HST image, a very faint extended source resembling a passive spiral galaxy is seen at a position consistent with this source.  This optical source is thus most likely an unrelated galaxy.
\end{itemize}

In summary, the simultaneous data obtained from the OM reveal only one possible association with knot 2. The SDSS catalogue has one point-like source near knot 2 (consistent with the OM source) and one extended source near knot 3. Their optical properties imply that the first one is as a star and the second one a galaxy. We estimated the probability of chance associations between the XMM sources and SDSS sources within a 2 arcsec radius to be 48\% (based on the source density of SDSS in the XMM field). This probability is consistent with finding unrelated SDSS sources near two out of the three knots.
The different nature of the SDSS optical sources, in contrast to the very similar X-ray spectra of the knots, suggest that these are very unlikely to be true associations.

\subsubsection{Starburst wind scenario}
Superwinds are often observed in nearby starburst galaxies. They display complex, non-aligned morphologies rather than collimated structures (e.g.
NGC253 \citep{pietsch2001,strickland2000} and NGC3079 \citep{pietsch1998}). In the starburst NGC2782, \cite{bravo} found a bubble structure with spectral properties consistent with the scenario of a super-wind observed in X-ray, radio and $H_{\alpha}$. 
While the presence of a starburst in NGC6217 is confirmed by the spectral analysis of its central core, the alignment of the knots with the core points towards the hypothesis of a jet injected by an AGN.  A strong argument to exclude the superwind scenario is also found in the spectral analysis, since the X-ray spectra of starburst winds are typically softer than those of the nuclei (for example, \citealt{strickland2000}).
Here we find that the knots have spectra with a photon index of  $\sim1.7$, significantly harder than the central core, as found also in \citep{pietsch2001} using the broad band X-ray colours.
This provides an argument against the starburst wind scenario. 

\subsubsection{Jet scenario}
The following discussion will focus on the last hypothesis: that the three knots are part of a jet emitted by the nucleus of NGC6217.
 
This scenario is supported by the similarities in the X-rays between the knots.
The spectral analysis shows that the three knots have power laws with a similar photon index of $\sim1.7$ within an error of  29\%. The joint fitting finds a photon index of 1.7, reducing the uncertainty to 15 \%. This value is consistent with other known jets in the literature (e.g. \citealt{worrall2016}).
 While the knot spectra can be fitted with a thermal component almost equally well, the high temperature of 5 keV makes it difficult to classify the knots as usual astrophysical sources. Thus, even if the thermal emission cannot be ruled out, the high temperature leaves the non thermal emission as the most probable scenario. Non-thermal emission can be originated either in synchrotron emission in the jet or Inverse Compton scattering, as discussed in \cite{worrall2009} and references therein. 

The discrete jet structure is consistent with several examples in the literature
and it is found as a common feature of jets (e.g. \citealt{massaro2011} and references therein). For NGC6217, the knot structure has been proposed on the basis of ROSAT data: the improvement of angular resolution between ROSAT PSPC and ROSAT HRI revealed the knot structure for the first time. Higher resolution is thus promising to better resolve the knots, and will also offer the interesting opportunity to detect any projected motion of the knots. 
For example, M81 has a jet with a discrete structure \citep{king2016}, for which an apparent velocity of the knots of 0.5$c$ was detected in radio observations.   The current XMM data do not allow us to make such an estimate: $\sim$10 years after the ROSAT observations any movement of the knots would still be hidden within a 10'' angular dispersion introduced by the X-ray instruments.

The jet is interestingly one-sided. Jets are expected to emerge from each face of the accretion disk. If the system is observed from a favourable direction, relativistic beaming will increase the apparent brightness of the approaching jet, increasing the brightness ratio between the jet and the counter jet. This explains the high percentage of one-sided jets; for example, they have been found in 94\% of the AGN surveyed with VLBA by \cite{lister2008}. We estimate that the jet in the SW direction is 8 times more luminous than in the opposite direction. This value has to be interpreted as a lower limit since the counter jet flux is taken as the background level. The ratio $>8$ is consistent with many jets studied with Chandra (e.g. \citealt{sambruna2004}).   Considering that the jet is likely beamed, its apparent length might be more extended that it actually is if the inclination is low.    

In NGC6217, the core and the jet luminosity are  7.2 $\times 10^{39}\ \rm{erg}$ $\rm{s^{-1}}$ and 5 $\times 10^{38}\ \rm{erg}$ $\rm{s^{-1}}$,  respectively, and the jet is extended over 15~kpc.
Several jets from galaxies hosting AGN at such low luminosities are known to be less extended on the kpc scale. For example, M87, with a luminosity of $10^{40}\  \rm{erg\ s^{-1}}$ has a jet extended to 1 kpc \citep{bohringer2001}.
On the other hand, objects with luminosities as high as 10$^{44}$ erg $\rm s^{-1}$ typically have larger jets, up to 100~kpc (e.g., \citealt{simionescu2016}).
However, there is no simple relation between jet power and length, since these properties are also affected by relativistic beaming and the interaction with the environment \citep{hardcastle2013}.

In radiatively efficient (standard) AGN, jet power and radiative output are broadly correlated \citep{rawlings1991,mingo2016}. This, however, is not the case for radiatively inefficient (ADAF) AGN, where most of the accretion energy is channelled into the jet \citep{narayan1995}.
These sources lack the accretion signatures of standard AGN, so the jets are often the only clear observational evidence for the presence of actively accreting SMBHs. It is therefore particularly interesting to assess the accretion mode of the active nucleus in NGC6217 and see how it compares to other low luminosity AGN at low redshift (see Sect. \ref{nucleus_discussion}).

Extragalactic jets are typically found in elliptical galaxies, with a relatively small number of examples of jets from spiral galaxies confirmed so far  (e.g. \citealt{ledlow1998}, \citealt{hota+2011}, \citealt{kharb2014}, \citealt{bagchi2014}, \citealt{kaviraj2015},  \citealt{singh2015}, \citealt{king2016},   \citealt{ledlow2001}, \citealt{croston2008}, \citealt{mingo2011}, \citealt{irwin2017}). NGC6217 is thus an excellent candidate for extending our knowledge of how the host and AGN influence the jet properties. 

\subsection{The nucleus of NGC6217}\label{nucleus_discussion}
NGC6217 has been optically classified as a LINER. Since there is evidence that LINERs exhibit AGN properties (e.g. \citealt{gonzalez2008}), it is interesting to view them as low-luminosity Seyfert nuclei. The variability is a distinctive property of AGN \citep{peterson1997}, and it offers compelling evidence for the presence of SMBHs. For this reason, we have performed a spectral analysis of the core of NGC6217 in each of the four XMM observations to search for any X-ray variability as a signal of an AGN.
The fluxes estimated from the individual spectral fits of the XMM observations are compatible within the uncertainties (section \ref{spectra}). This means that an AGN is unlikely to dominate the X-ray spectrum. This conclusion is also supported by the X-ray spectrum, which is well described by a thermal plasma as discussed below.

Our analysis of the XMM spectra of NGC6217 showed that the spectrum can be equally well described by a mixed starburst - AGN model and a 'pure' starburst model. In all models the absorption is moderate, being only an upper limit of $<0.27\times10^{22}\ \rm{cm^{-2}}$ in the starburst-AGN scenario and constrained to $0.58\pm0.08\times10^{22}\ \rm{cm^{-2}}$ in the pure starburst model. As explained in \cite{caccianiga2007}, if the absorption has column density  $\rm{N_H}\geq4\times10^{21}\ \rm{cm^{-2}}$ the ISM is not enough to explain the obscuration of the optical lines and a torus is needed. So, column densities higher than this threshold have been adopted to classify AGN as type 2 based on their X-ray spectra.
In the XMM fits of NGC6217 the absorption is below or around this threshold, so it is consistent with a non-obscured AGN, at odds with the lack of broad optical lines in the nucleus of NGC6217 reported in \cite{kennicutt1992}. The mismatch between the two classifications can be explained if there is variability in the absorber column density along the line of sight. This is reasonable in the time interval between the optical and X-ray observations: significant variability in the absorption is broadly reported in the literature of AGN from low redshift (e.g., \citealt{puccetti2004}) to high redshift (e.g. \citealt{shemmer2014} and references therein).

The starburst-AGN scenario is similar to the model used in \cite{pietsch2001}, which found an AGN powerlaw (with $\Gamma\sim1.8$) and a thermal component with a temperature of 0.5 keV. In our analysis we find a temperature of the thermal plasma similar to \cite{pietsch2001} (0.6 keV) but a steeper powerlaw  (with $\Gamma\sim3$).
The different photon index found in the XMM observations is likely due to the fact that the ROSAT data extends to lower energies. 

For the fits in the pure starburst scenario, the two temperatures of 0.3 keV and  0.67 keV are typical of a spiral galaxy disk and a nuclear starburst, respectively, as found in the analysis of M83 by \cite{soria2001}.
In this model, the emission from the spiral arms in NGC6217 is non-negligible, since it contributes 43\% to the full observed flux.
In the starburst - AGN scenario, the observed flux is 1.6$\times10^{-13}\ \rm{erg}$  $\rm{cm^{-2} s^{-1}}$, out of which the AGN contributes  $46 \%$.


To further investigate if the core of NGC6217 harbours an AGN we searched for counterparts in the OM data and found an association in the observations from 2002  and 2007.  The $u$ band magnitudes from the first to the last observations are 13.70$\pm0.02$, 13.61$\pm0.01$ and 13.80$\pm0.01$.
The non-variability of the nucleus of NGC6217 in the optical band, at odds with the widely found optical variability of AGN nuclei, is consistent with the non-variable X-ray emission and suggests that NGC6217 lacks standard accretion activity in its core.
The resulting X-ray to optical flux ratio of -2.7 is consistent with the one presented by \cite{pietsch2001}. This is below the value typically found in AGN  \citep{mainieri2002}.

To place constraints on the accretion mode in the nucleus of NGC6217, we computed the black hole mass and the bolometric luminosity. The black hole mass was estimated using the \cite{ferrarese2000} scaling law with the observed stellar velocity dispersion of 70$\pm$10 km/s  (\citealt{ho2009}), resulting in a value of $M_{\rm{BH}} = 8\times10^6$ $M_{\sun}$. The bolometric luminosity was computed from the \cite{marconi2004} relation, using the observed luminosity in the 0.5-2 keV rest-frame band of  6.8 $\times 10^{39}\ \rm{erg\ s^{-1}}$ (from the AGN component only).
We computed the Eddington ratio as $\rm{log} ( L_{\rm{bol}}/L_{\rm{edd}}$), where $L_{\rm{bol}}$ is the bolometric luminosity and $L_{\rm{edd}}$ is the Eddington luminosity computed from the black hole mass. The resulting Eddington ratio for NGC6217 is  -4.09, consistent with the values reported in the literature for well known LL AGN such as M81 and NGC4579 \citep{quataert1999}. These have been interpreted as hosting ADAFs (\citealt{narayan1995} and references therein). For these sources \cite{quataert1999} reported a configuration of a standard disk plus an ADAF similar to the model proposed for X-ray binaries \citep{esin}. The transition radius between the standard disk and the ADAF depends on the Eddington ratios \citep{yuan2004} and for a value of  -4.09 it is 100$R_s$, as in M81 and NGC4579 \citep{quataert1999}. The lack of reflection features (especially iron lines) in the spectrum of the core of NGC6217 is expected in a scenario with a weak primary power law and an ADAF. 

The bolometric luminosity of NGC6217, together with the luminosity at 151 MHz (570 mJy as inferred from the flux value of 96 mJy from the NVSS 1.4 GHz map) can be directly compared with values reported for similar sources in the MIXR (Mid Infrared, X-ray and Radio) sample of \cite{mingo2016}. NGC6217 is located in the area occupied by LINER and LERG (Low Excitation Radio Galaxies) of the MIXR sample in the plane formed by $\log{L_{bol}}$ and $\log{Q}$ (where $L_{bol}$ is the bolometric luminosity in Watt and Q is the jet output as defined in equation (10) of \cite{mingo2016}, see Fig. 27 of \citealt{mingo2016}).
In particular, the value of $\frac {\log{L_{bol}}} { \log{Q} }\sim 0.92$  is consistent with the wide range of values found for LINER and LERG \citep{mingo2016}. 
High star-formation rates and jets are thought to be related with frequent interactions between galaxies and their nearby companions. In particular, interactions result in dense environments   that influence star formation.
Moreover, there is evidence for an interplay between the timing of mergers, AGN activity and star formation, and an ongoing discussion on whether radio AGN activity would trigger or quench star formation \citep{wild2010}.
For these reasons it is interesting to consider the cluster environment of NGC6217. The HyperLeda catalogue includes NGC6217 and gives information on the companions of the galaxy within a radius of $4''$ \citep{makarov2014}. It lists eight associated galaxies in a galaxy group with a velocity dispersion of 374 km/s. The galaxy is in an environment where interactions can play a role in accelerating its star formation processes.

\section{Summary and future perspectives}

NGC6217 is a spiral galaxy with a LINER nucleus. HST images highlight
a complex nuclear structure with a double core. The first X-ray
analysis based on ROSAT data unveiled a possible jet formed by three knots in
the soft X-rays \citep{pietsch2001}. However, the ROSAT spectra of the knots
were not analysed spectroscopically due to the low count statistics. This paper presents an X-ray spectral analysis of the
knots and the core, as well as a study of the images, exploiting XMM
data. The XMM images confirm the jet structure formed by three knots
aligned with the nucleus. The nucleus and the knots emit the majority
of their radiation in the soft X-rays. Our spectral analysis confirms
that the knots have spectra consistent between them, providing support for the jet
hypothesis. The spectra of the knots have a harder photon index than the core, and this represents an argument against the hypothesis that they are part of a starburst wind.

The compact nucleus of NGC6217 has a flux of
$1.6\times10^{-13}\ \rm{erg\ s^{-1}}$ and the spectrum is equally well fitted by a pure starburst model (with contributions from the nucleus and spirral arms) and a model with a nuclear starburst and an AGN. In the latter model the AGN makes up  46\% of the total flux, which sets an upper limit on the AGN contribution.  It is 
important to find a robust probe of the AGN at the centre of this galaxy because it would provide independent support for the jet hypothesis, since a pure starburst cannot produce such an aligned
structure.  We estimated the Eddington Ratio of the putative AGN, using the black hole mass estimated from the stellar velocity dispersion together with the X-ray luminosity. This gives a value of    $\rm{log} ( L_{\rm{bol}}/L_{\rm{edd}})= -4.09$, which is in agreement with LLAGN known to be powered by ADAFs.

Observations with a higher spatial resolution are needed in order to better constrain the nature of the core of NGC6217 and to solve the puzzle of its possible jet. In particular, with Chandra resolution it will be possible to disentangle the compact nucleus from the diffuse emission of the galaxy, place constraints on the X-ray emission from the two cores seen in the HST image, and  determine if the knots are spatially extended. 
 Additionally, new radio data capable to resolve the knot structure observed in the X-rays will be necessary to confirm the presence of a radio jet. 
Summarising, while the currently available data from XMM provide support for the jet scenario, a further improvement in angular resolution is necessary to make significant progress in understanding the nature of the source. 

\section*{Acknowledgements}
We acknowledge the referee for his/her helpful suggestions that have improved the paper significantly. 
This work was supported by the Swedish National Space Board and the Knut \& Alice Wallenberg Foundation.
SN acknowledges financial support by Wenner-Gren foundation (Stockholm,Sweden).




\bibliographystyle{mnras}
\bibliography{bibliography} 



\appendix


\bsp	
\label{lastpage}
\end{document}